\documentclass[10pt,onecolumn]{aastex631}

\usepackage{natbib}
\usepackage{graphicx}
\usepackage{subfigure}
\usepackage{amsmath,amssymb}
\newcommand{\beginsupplement}{%
        \setcounter{figure}{0}
        \renewcommand{\thefigure}{S\arabic{figure}}%
        \renewcommand{\theHfigure}{S\arabic{figure}}%
     }

\defcitealias{Xing2024}{Paper I}

\shorttitle{Initiation Route of CMEs: II. Role of Filament Mass}
\shortauthors{Xing et al.}

\begin{document}

\title{Initiation Route of Coronal Mass Ejections: II. The Role of Filament Mass}

\correspondingauthor{Chen Xing, Xin Cheng}
\email{chenxing@nju.edu.cn, xincheng@nju.edu.cn}

\author{Chen Xing}
\affiliation{School of Astronomy and Space Science, Nanjing University, Nanjing, China}
\affiliation{State Key Laboratory of Space Weather, Chinese Academy of Sciences, Beijing, China}

\author{Xin Cheng}
\affiliation{School of Astronomy and Space Science, Nanjing University, Nanjing, China}

\author{Guillaume Aulanier}
\affiliation{Sorbonne Université, Observatoire de Paris-PSL, École Polytechnique, IP Paris, CNRS, Laboratoire de Physique des Plasmas, Paris, France}
\affiliation{Rosseland Centre for Solar Physics (RoCS), Institute of Theoretical Astrophysics, Universitetet i Olso, Oslo, Norway}

\author{Mingde Ding}
\affiliation{School of Astronomy and Space Science, Nanjing University, Nanjing, China}

\begin{abstract}
The thorough understanding on the initiation of coronal mass ejections (CMEs), which is manifested as a slow rise of pre-eruptive structures before the impulsive ejection in kinematics, is the key for forecasting the solar eruptions. In our previous work, we showed that the slow rise of a hot flux rope with coronal mass density is caused by the moderate magnetic reconnection occurring in the hyperbolic flux tube (HFT) combined with the torus instability. However, it remains unclear how the initiation process varies when a filament is present in the pre-eruptive flux rope. In this work, we reveal the complete initiation route of a CME containing filament mass with a state-of-the-art full-magnetohydrodynamics simulation. The comprehensive analyses show that the filament mass has an important impact on the CME initiation through triggering and driving the slow rise of flux rope with its drainage, besides the contributions of HFT reconnection and torus instability. Finally, in combination with our previous work, we propose that the enhanced drainage of filament mass and various features related to the HFT reconnection, such as, the split of pre-eruptive structure and the pre-flare loops and X-ray emissions, can serve as the precursors of CME initiation in observations.
\end{abstract}

\keywords{Sun: corona, Sun: coronal mass ejections (CMEs), Sun: filaments, Sun: flares}

\section{Introduction}\label{sec1}
Coronal mass ejections (CMEs) are one of the most violent solar eruptions, and they could have serious impacts on human high-tech activities, e.g., aerospace activity, satellite communications, and power transmission when arriving at the Earth \citep{Gosling1993}. Therefore, understanding and forecasting CMEs has been a hot topic in solar physics for the last few decades, and it even makes an important contribution to exploring the habitable planets through applications in the field of stellar coronal mass ejections \citep{Veronig2021}.

CMEs originate from the eruption of a pre-eruptive structure which usually appears as filaments/prominences or hot channels \citep{Cheng2011}. The filament refers to the dark structure in H$\alpha$ and EUV passbands on disk, while the prominence appears as a bright structure above the solar limb \citep{Mackay2010}. The filament and prominence are actually the same structure appearing at different positions (on disk/above limb), so we will not distinguish between them in the following. The hot channel is a bright structure at high-temperature passbands, i.e., 131 \AA\ and 94 \AA\ \citep{Cheng2011,Zhang2012}. Generally, it is considered that the magnetic configuration of CMEs and most of their pre-eruptive structures is a flux rope \citep{Aulanier1998,Dere1999,Guo2010,Zhang2012,Ouyang2017}, defined as a coherent structure composed of field lines winding around a central axis \citep{Patsourakos2020}. Importantly, the flux rope can support the filament, with its Lorentz force balancing the gravity force of the dense and cold plasmas of filament  \citep{Hirayama1985,Gibson2001}.

Understanding the kinematics of CMEs and their pre-eruptive structures is the first step for developing the strategy of space weather prediction. The observational studies show that the kinematics of a CME event is usually composed of four phases. The first phase is the quasi-static phase in a long period, e.g., several hours or days, before the eruption. In this phase, the pre-eruptive structure rises quasi-statically with a velocity less than 1 km s$^{-1}$ and an acceleration close to zero \citep{Sterling2004,Schmieder2008,Xing2018}. The second phase, i.e., the slow rise phase, is also before the eruption but just following after the quasi-static phase \citep{Sterling2011,Vemareddy2017,Cheng2020}. The slow rise phase lasts for several to tens of minutes, in which the pre-eruptive structures rise with a velocity of tens of km s$^{-1}$ and an acceleration of tens of m s$^{-2}$ \citep{Cheng2020,Cheng2023}. The third and the fourth phases are the impulsive acceleration phase and the propagation phase, respectively, in the former of which the velocity of CMEs is sharply accelerated to hundreds or even thousands of km s$^{-1}$ with an acceleration of hundreds of m s$^{-2}$ in tens to hundreds of minutes \citep{Zhang2006,McCauley2015}. Therefore, the slow rise phase represents the critical transition from the quasi-static state of pre-eruptive structures to the erupting state of CMEs, i.e., the so-called initiation process of CMEs.

The mechanisms of slow rise phase attract great interests as they are the key to understand how the pre-eruptive structures are initiated towards the eruption. In the past, it was generally considered that the slow rise is triggered and driven by one single mechanism, although there are different views on the specific mechanism \citep{Zhang2001,Savcheva2012}. On the one hand, some studies consider that the slow rise is caused by the eruption-trigger mechanisms at their early stages of development \citep{Zhang2001,Zhang2006}. The candidate mechanisms in this framework include the torus instability \citep{Kliem2006}, the breakout reconnection \citep{Antiochos1999}, and the tether-cutting reconnection \citep{Jiang2021}. On the other hand, some works argue that it is the moderate magnetic reconnection before the start of eruption-trigger mechanisms that leads to the slow rise by building up the flux rope and breaking the balance between the upward and downward Lorentz forces acting on the pre-eruptive structure \citep{Cheng2023}. However, since these two mechanisms (eruption-trigger mechanism and moderate magnetic reconnection) occur in different periods, these two interpretations, if both correct, contradict with each other and challenge the idea that the slow rise is due to one single mechanism.

One possible solution of such a paradox is first proposed by \citet[][hereafter referred to as \citetalias{Xing2024}]{Xing2024}, which reveals that the CME initiation is actually a multi-physics coupled process. For a hot flux rope with a coronal mass density (corresponding to the hot channel in observations), its slow rise is first triggered and driven by the moderate magnetic reconnection in the hyperbolic flux tube (HFT; \citeauthor{Titov2002}, \citeyear{Titov2002}) below the flux rope. Later, the torus instability starts during the initiation process, and the early development of it continues to drive the slow rise of flux rope together with the HFT reconnection. The slow rise ends as the fast magnetic reconnection starts, the latter of which triggers and drives the following impulsive acceleration of CME together with the torus instability \citepalias{Xing2024}. In addition, a detailed analysis on the dynamics shows that the acceleration of such a hot-channel-like flux rope during its slow rise is mainly contributed by the Lorentz force while rarely by the gravity force \citepalias{Xing2024}.

Nevertheless, there are still many questions about the mechanisms of CME initiation. As shown in \citet{Cheng2020}, not only the hot channels but also the filaments experience a slow rise phase before the eruption. A natural question is: what is the initiation route of CMEs containing filament mass? More essentially, what and how does the filament mass play a role in the slow rise of flux rope? These questions are not answered in \citetalias{Xing2024} as the modeled flux rope there does not carry dense plasmas. Therefore, to answer these questions, we carry out a new full-magnetohydrodynamics (full-MHD) CME simulation with the code MPI-AMRVAC \citep{Xia2018}. The comprehensive analyses reveal that multiple physics are involved in the initiation of CMEs containing filament mass. The initiation mechanisms include not only the HFT reconnection and the torus instability revealed in \citetalias{Xing2024}, but also the drainage of filament mass which plays an important role in triggering and driving the slow rise of flux rope. The overview and detailed analyses of the simulation are described in Section \ref{sec2}, and the complete initiation route of modeled CME is clarified in Section \ref{sec3}. The summary and discussions on the CME initiation are in Section \ref{sec4}, which is followed by a description of the setups of simulation in Appendix \ref{app1} and a supplementary dynamic analysis of simulation in Appendix \ref{app2}.

\section{Evolution of Modeled CME Event}\label{sec2}

\subsection{Overview of Modeled CME Event}

\begin{figure*}
\begin{interactive}{animation}{Movie_S1.mp4}
\centering
\includegraphics[width=\hsize]{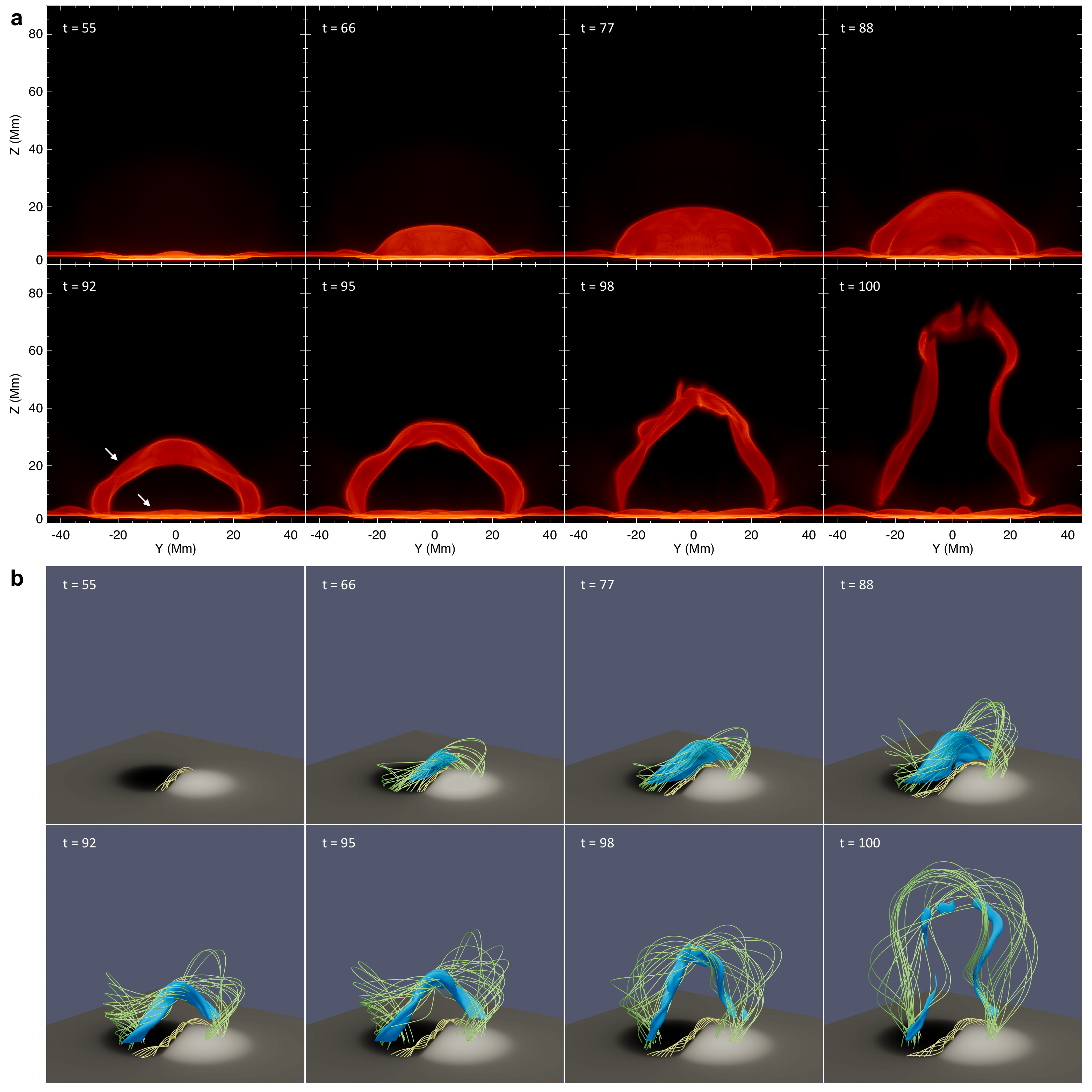}
\end{interactive}
\caption{\textbf{Overview of the evolution of modeled CME event.} (a) The synthetic extreme ultraviolet (EUV) images of simulation at 304 \AA\ from the side view. The two arrows point to the upper and the lower filaments at $t=92$, respectively. (b) An overview of the flux rope evolution. The cell-center bottom surface shows the distribution of vertical magnetic field $B_z$. Before $t=88$, the green and the yellow field lines represent the upper and the lower parts of the flux rope, respectively; since $t=88$, the green and the yellow field lines represent the upper and the lower flux ropes, respectively. The blue surface (isosurface of mass density equal to 5) roughly outlines the filament above the plane $z=0.6$. An animation of panel a is available, showing the evolution of filament from $t=55$ to $t=101$ in the simulation. The duration of the animation is 16 s.}
\label{fig1}
\end{figure*}

We study the initiation of a CME containing filament mass by performing an observationally-inspired full-MHD simulation, the setups of which are described in Appendix \ref{app1}. Fig \ref{fig1} and Movie 1 give an overview of the evolution of modeled CME event. Driven by the converging flows towards the polarity inversion line (PIL), a pre-eruptive flux rope is first formed by the magnetic reconnection between the sheared arcades in the flux cancellation framework (e.g., $t=66$ and $t=77$ in Fig \ref{fig1}b), and a filament is gradually formed at the same time (Fig \ref{fig1}a). Shortly before the eruption onset, the filament/flux rope is split into two parts (e.g., $t=88$ and $t=92$ in Fig \ref{fig1}; see Section \ref{split} for more details of this process), matching well with the double-decker filament/flux rope configuration in observations \citep{Liu2012,Pan2021}. Later, the upper filament/flux rope erupts as a CME, while the lower part survives and lies below the post-flare loops (e.g., $t=98$ and $t=100$ in Fig \ref{fig1}).

The modeled filament is most likely to be formed as the flux rope lifts cold and dense plasmas from the chromosphere and transition region into the corona during its growth and rising, i.e., in the framework of the well-known levitation model \citep[see more details of levitation model in the review by][]{Karpen2015}. This process is evidenced by the quick influx of dense plasmas from the lower atmosphere into the corona during the filament formation (Movie 1 and Fig \ref{fig4}a). Such a view is also supported by a control simulation in which only the initial atmosphere is changed to reduce the mass density of chromosphere and transition region. Compared to the filament in this work, the filament in the control simulation has less plasmas which finally all fall back to the lower atmosphere once the eruption starts. This suggests that the modeled filaments are strongly influenced by the setups of lower atmosphere, so it is most likely that they originate directly from the lower atmosphere. In addition, it is interesting to note that the modeled filament is distributed not only in the middle section of flux rope but also along its legs (Fig \ref{fig1}b), the latter of which consists of the dense plasmas falling from the top of filament (Fig \ref{fig4}a-c). More details of the formation process of modeled filament will be studied in our upcoming work.

\subsection{Kinematics of CME Event}

\begin{figure*}
\centering
\includegraphics[width=\hsize]{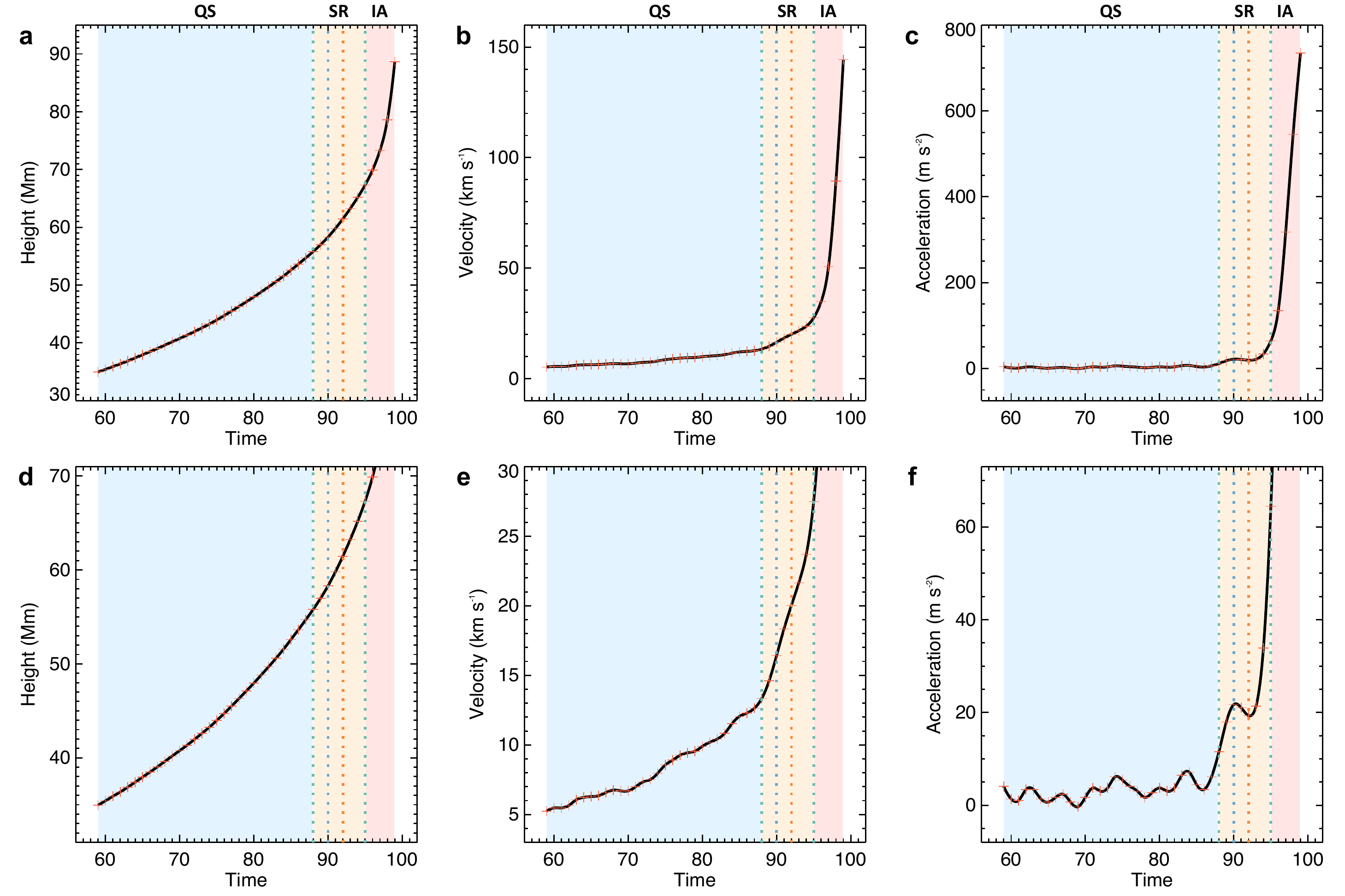}
\caption{\textbf{Kinematics of the CME event.} (a)-(c) The evolution of height, velocity, and acceleration of the eruptive flux rope during $59\le t\le99$. The blue, yellow, and red regions mark the quasi-static phase (QS), slow rise phase (SR), and impulsive acceleration phase (IA), respectively. The two green dashed lines mark the start and the end of slow rise phase. The blue and orange dashed lines mark a peak and a dip of the acceleration during the slow rise phase, respectively. (d)-(f) Zoom-in images of panels a-c.}
\label{fig2}
\end{figure*}

Fig \ref{fig2} shows the kinematics of the modeled CME event during $59\le t\le99$, which is derived by measuring the height of the apex of an overlying field line right above the eruptive flux rope. This overlying field line is traced from a fixed point at the center of the positive polarity and on the cell-center bottom surface (an horizontal layer at the cell center of the first layer of the physical simulation domain). The velocity at the fixed point is very small throughout the simulation and this overlying field line does not experience magnetic reconnection during $59\le t\le99$, which ensures the accurate measurement of kinematics. The kinematics after $t=99$ is not analyzed in this work, since the overlying field line is reconnected in this period and can no longer be used to estimate the kinematics.

It is obvious that the flux rope experiences an impulsive acceleration phase during $95<t\le99$ (Fig \ref{fig2}b,c). The evidences are (1) the velocity of flux rope has a significantly rapid increase after $t=95$ when compared to that before $t=95$ and (2) the acceleration of flux rope is larger than 100 m s$^{-2}$ (the empirical value for dividing the impulsive acceleration phase and the slow rise phase in observations) after $t=95$. The pre-eruptive stage of the CME event ($59\le t\le95$) is clearly composed of two phases (Fig \ref{fig2}f): the quasi-static phase in $59\le t<88$ when the flux rope acceleration is oscillating around a quite small value close to zero, and the slow rise phase in $88\le t\le95$ when the eruptive flux rope rises with an acceleration on the order of tens of m s$^{-2}$. The clear change in the evolution of rising velocity before and after $t=88$ (Fig \ref{fig2}e) also shows that the pre-eruptive stage consists of these two phases.

\subsection{Split of Filament and Flux rope}\label{split}

\begin{figure*}
\centering
\includegraphics[width=\hsize]{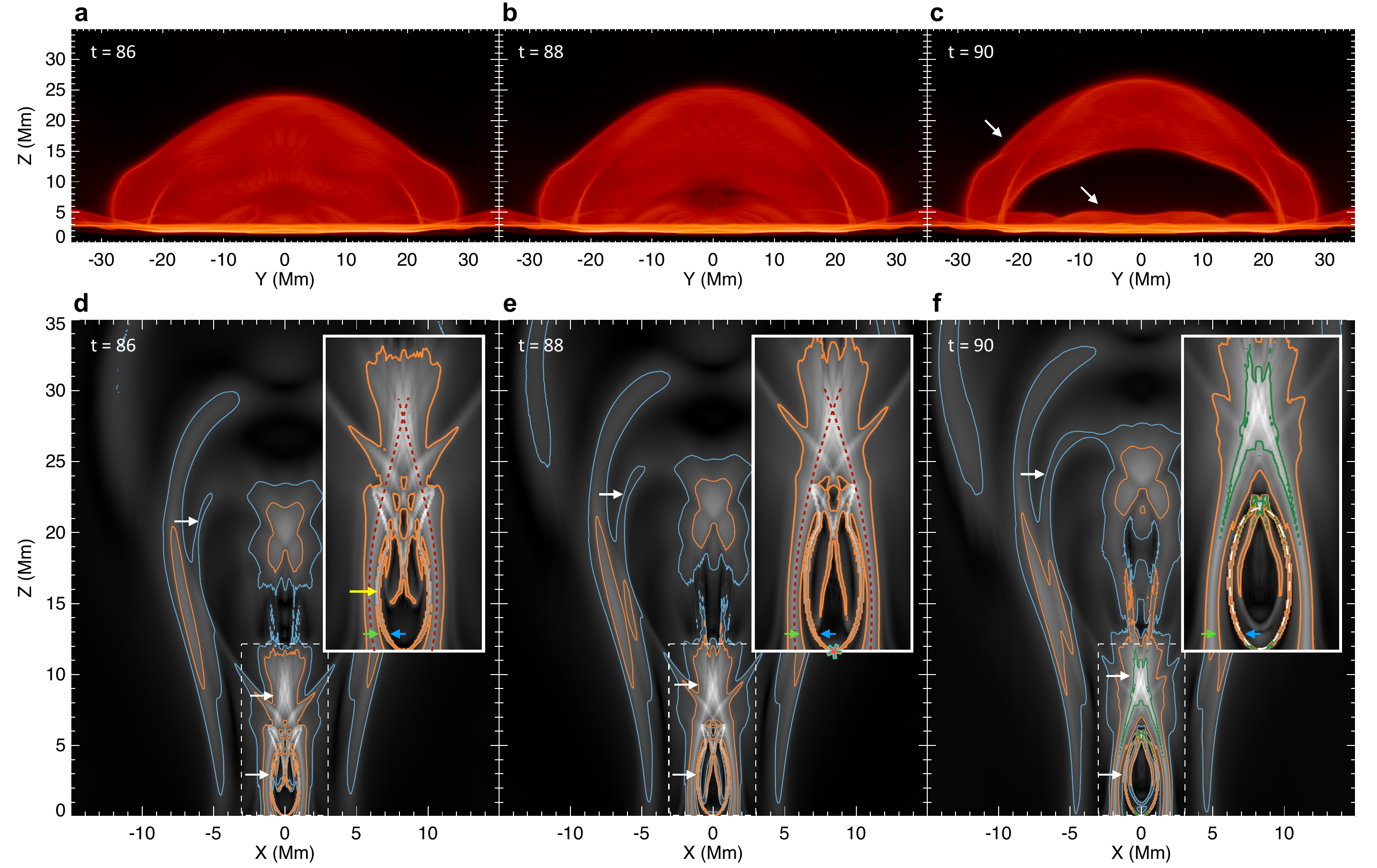}
\caption{\textbf{Evolutions of the filament and flux rope during their split.} (a)-(c) The synthetic EUV images of filaments at 304 \AA\ from the side view. The white arrows point to the upper and the lower filaments at $t=90$, respectively. (d)-(f) The squashing degree ($\log Q$) maps in the plane $y=0$ which is perpendicular to the upper flux rope axis. The blue, orange, and green curves are the contours of $\log Q=2$, $\log Q=3$, and $\log Q=5$, respectively. We use the contour of $\log Q=2$ to outline the upper flux rope, because its boundary is very diffuse. In panels d-f, the three white arrows from top to bottom point to the upper flux rope, the HFT, and the lower flux rope, respectively; the zoom-in image at the upper right corner of each panel shows the details of HFT and lower flux rope. The skeleton of HFT is marked by the red dashed lines at $t=86$ and $t=88$; at $t=90$, the core region of HFT is marked by the contour of $\log Q=5$. The lower flux rope at $t=90$ is roughly outlined by the white dashed oval. In panels d-f, the green arrow points to the boundary of HFT, while the blue arrow points to the boundary of lower flux rope. In panel d, the yellow arrow points to the interface of the boundaries of HFT and lower flux rope. The red asterisk in panel e marks the position of BP.}
\label{fig3}
\end{figure*}

The synthetic EUV images clearly show the split of filament shortly before the eruption (Movie 1). As shown in Fig \ref{fig1} and Fig \ref{fig3}a, there is only one filament for a long period before the eruption (e.g., $t=66$, $t=77$, and $t=86$). However, the 304 \AA\ emission at the central part of filament starts to decrease at $t=88$ (Fig \ref{fig3}b), indicating a reduction of the local plasmas. Later, the 304 \AA\ emission at the central part further decreases and a cavity appears there at $t=90$ (Fig \ref{fig3}c), marking that the original filament has been split into two parts at this moment.

We also study the split of filament from the point of view of the magnetic field connectivity. Fig \ref{fig3}d-f shows the distribution of squashing degree ($Q$) in the plane $y=0$ which is perpendicular to the upper flux rope axis. The quasi-separatrix layers \citep[QSLs;][]{Priest1995,Demoulin1996}, which correspond to the regions where $\log Q\gg2$ \citep{Titov2002}, outline the flux ropes and reconnection regions. One can find that there is an HFT located between the upper and lower flux ropes during $86\le t\le90$ (Fig \ref{fig3}d-f) and that the split of filament occurs exactly in the HFT (Fig \ref{fig3}).

The detailed analyses on the magnetic field connectivity reveals the split of flux rope accompanying the filament split. During $86\le t\le90$, the magnetic reconnection occurs in two regions, i.e., the bald patch \citep[BP, the position of which is marked by the red asterisk in Fig \ref{fig3}e;][]{Titov1993} below the lower flux rope and at the bottom surface, and the HFT between the upper and lower flux ropes (Fig \ref{fig3}d-f). The BP reconnection mainly contributes to the formation of lower flux rope. In contrast, the HFT reconnection builds up the upper flux rope, which means that the HFT is always connected with the upper flux rope. One can find that the boundaries of lower flux rope (pointed by the blue arrow in Fig \ref{fig3}d) are connected with the boundaries of HFT (pointed by the green arrow in Fig \ref{fig3}d) at many places at $t=86$. The yellow arrow in Fig \ref{fig3}d exactly marks one of the interfaces of the boundaries of lower flux rope and HFT at this moment. Therefore, we consider that the upper and lower flux ropes are connected at $t=86$, with the HFT as a linker. In other words, these two flux ropes can be regarded as the substructures belonging to one single flux rope at this moment. Later (during $86\le t\le90$), as the HFT reconnection is enhanced (Fig \ref{fig5}b,c), the HFT grows with its core region rising and its two legs separated far from each other (Fig \ref{fig3}d-f); in comparison, the growth of lower flux rope is relatively slow in the same period (Fig \ref{fig3}d-f). As a result, the lower flux rope and the HFT are separated on most of their original interfaces at $t=88$ (Fig \ref{fig3}e), and they are completely separated from each other at $t=90$ (Fig \ref{fig3}f). In other words, the upper and lower flux ropes can no longer be topologically considered as belonging to one single flux rope at $t=90$, that is the so-called split of flux rope.

The split process of flux rope in this simulation seems to be in the framework of split model revealed in \citet{Gibson2006}, the latter of which shows that the flux rope is split into two parts by the magnetic reconnection occurring in a central vertical current sheet in the corona. However, our split process still shows some differences. In \citet{Gibson2006}, the magnetic reconnection in the central vertical current sheet occurs between two field lines lying in the bald-patch separatrix surface, forming the upper and lower flux ropes simultaneously and thus leading to the split of flux rope. In contrast, in our simulation, the upper and the lower flux ropes are formed by the magnetic reconnection occurring in the HFT in corona and the BP at photosphere, respectively. The field lines involved in the HFT and the BP reconnection are two different groups of field lines, respectively, as the former is anchored at the feet of HFT while the latter is anchored at the BP (Fig \ref{fig3}d-f). The split of flux rope (i.e., the separation of two flux ropes) is due to that the HFT is built up more quickly than the formation of lower flux rope (Fig \ref{fig3}d-f).

In short, we conclude that the split of filament and flux rope starts at $t\approx88$ and is well developed at $t\approx90$, after which the upper filament/flux rope soon erupts as a CME.

\subsection{Drainage of Filament Mass}\label{drainage}

\begin{figure*}
\centering
\includegraphics[width=\hsize]{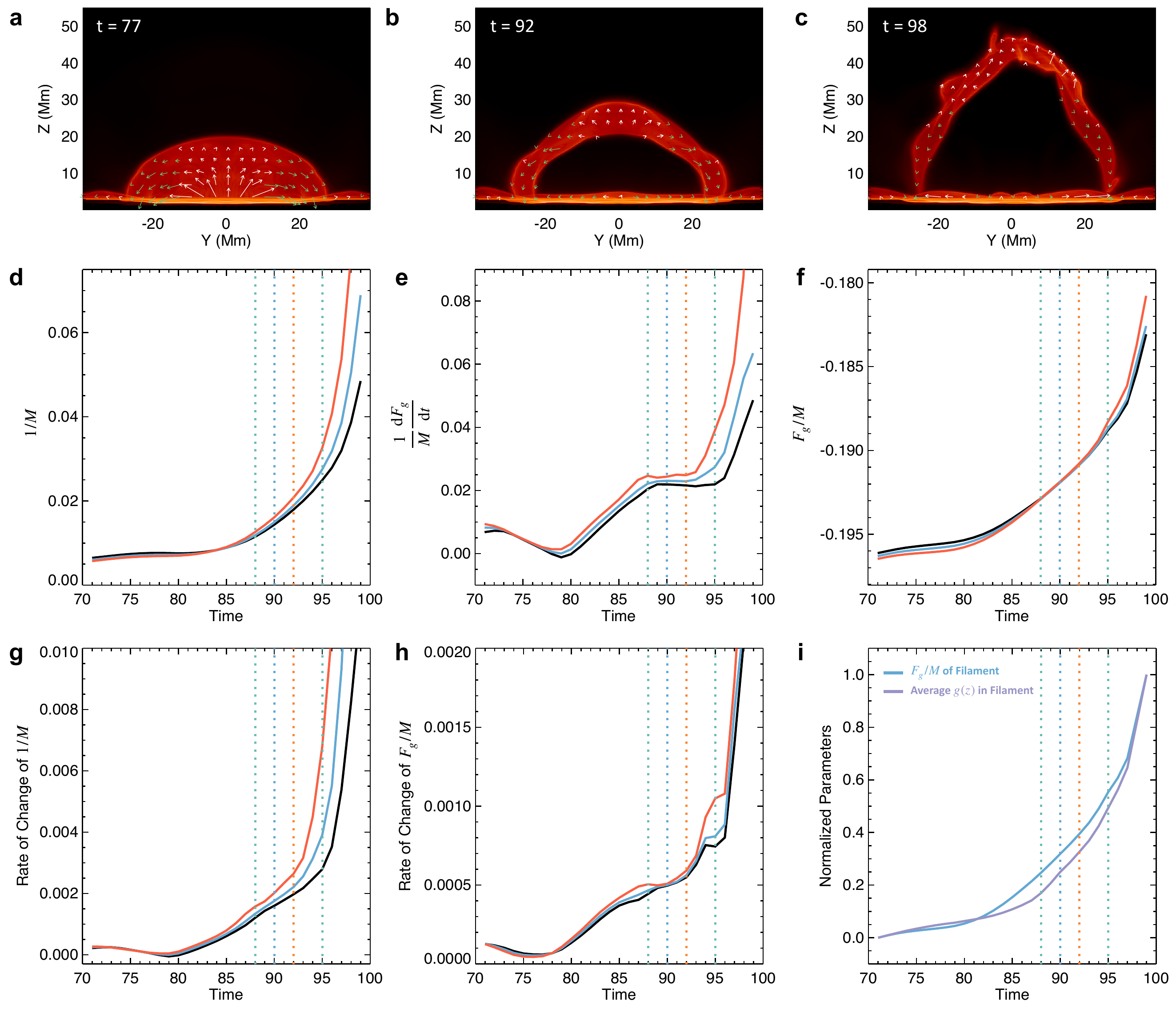}
\caption{\textbf{Drainage of filament mass and its effects on the kinematics.} (a)-(c) The background is the synthetic EUV images at 304 \AA\ from the side view. The arrows represent the integral of the momentum $\rho v_y\boldsymbol{e_y}+\rho v_z\boldsymbol{e_z}$ along the $x$ direction in the high-mass-density region ($\rho\ge5$). The arrows are shown only where the synthetic 304 \AA\ emission is strong, with the white (green) arrows marking the upward (downward) flows. The length scales of arrows in panels a-c are the same. (d) The evolution of the reciprocal of the mass of upper filament/flux rope, $1/M$. The black, blue, and red curves show the estimated parameters $1/M$ with three thresholds of mass density ($\rho_1=2.5$, $\rho_2=5$, and $\rho_3=10$) to identify the filament, respectively. (e) The evolution of the parameter $(\mathrm{d}F_g/\mathrm{d}t)/M$ which represents the rate of change of the flux rope acceleration contributed by the force imbalance due to filament drainage. (f) The evolution of the gravitational acceleration of upper filament/flux rope, $F_g/M$. (g) The rate of change of the parameter $1/M$. (h) The rate of change of the parameter $F_g/M$. (i) The evolutions of the parameter $F_g/M$ (blue curve, same with the blue curve in panel f) and the average of gravitational acceleration coefficient, $g(z)$, in the upper filament (purple curve). These two parameters are for the filament identified with the mass density threshold $\rho_2=5$, and their ranges are both normalized into [0,1] with their first data set to 0. The black, blue, and red curves in each of panels e-h have the same meanings as those in panel d but for the parameter in each panel, respectively. The green, blue, and orange dashed lines in panels d-i have the same meanings as those in Fig \ref{fig2}.}
\label{fig4}
\end{figure*}

The dense plasmas of the eruptive filament are drained before and during the eruption, as shown by the downward flows in the filament (Fig \ref{fig4}a-c). To evaluate the effect of the drainage of filament mass on the kinematics of CME event, we first identify the eruptive filament to calculate its mass and the gravity force acting on it. During $89\le t\le99$, considering that the upper and lower filaments are clearly separated from each other, the eruptive/upper filament is identified as the summation of dense plasmas above a fixed horizontal plane. The dense plasmas of filament are determined as those with a mass density ($\rho$) exceeding a threshold value, and here we set three threshold values, i.e., $\rho_1=2.5$, $\rho_2=5$, and $\rho_3=10$. The horizontal plane is as low as possible but still slightly higher than the lower filament. Then, one eruptive filament is identified in the case of each threshold value of mass density. During $71\le t\le88$, the upper and lower parts of filament have not yet been (completely) split from each other. Nevertheless, we still identify the upper part of filament by the method mentioned above, assuming that the lower part of filament changes little. The mass of eruptive filament ($M$) and the gravity force acting on it ($F_g$; $F_g<0$) are then derived by integrating the mass density ($\rho$) and the product of mass density and gravitational acceleration coefficient ($f_g=\rho g(z)$; $g(z)<0$, as shown in Equation \ref{eq8}) in the eruptive filament, respectively. In the following, we assume that the mass of eruptive filament and the gravity force acting on it are substitutes for those of eruptive flux rope, considering that the strong gravity force distributed within the eruptive flux rope is almost entirely concentrated in the eruptive filament (see Appendix \ref{app2} and Fig \ref{figA1}a).

Fig \ref{figA1} shows the distributions of various forces within the eruptive flux rope. As driven by these forces including the gravity force, the acceleration of eruptive flux rope is given by:
\begin{equation}
a=\frac{F_n}{M}=\frac{F_o+F_g}{M}=\frac{F_o}{M}-\frac{|F_g|}{M},
\end{equation}
where $F_n$, $F_g$, and $F_o$, derived by integrating $f_n$, $f_g$, and $f_o$ (see Appendix \ref{app2} and Fig \ref{figA1}) in the volume of eruptive flux rope, represent the $z$-component of net force, the gravity force, and the $z$-component of the net of other forces in addition to the gravity force acting on the flux rope, respectively. Therefore, we can describe the effects of the drainage of filament mass on the acceleration of eruptive flux rope from three aspects:
\begin{itemize}
\item[1.] As the drainage decreases the mass of eruptive flux rope ($M$), the acceleration of flux rope will increase even if the net force ($F_n$) keeps unchanged.
\item[2.] During the drainage, the absolute value of the gravity force acting on the eruptive flux rope ($|F_g|$) will decrease with the reduction of its mass. This will lead to or enhance a force imbalance between the net of other forces ($F_o$) and the gravity force ($F_g$) and thus results in a larger net force ($F_n$) even if the net of other forces ($F_o$) keeps unchanged, which will magnify the flux rope acceleration.
\item[3.] During the drainage, the absolute value of the acceleration of eruptive flux rope provided by its gravity force ($|F_g|/M$; hereafter referred to as the absolute value of gravitational acceleration) will decrease if the absolute value of the gravity force acting on the flux rope is reduced faster than the mass of flux rope. Such a decrease of the parameter $|F_g|/M$ will cause the increase of flux rope acceleration even if the change in the parameter $F_o/M$ is not taken into account.
\end{itemize}

Accordingly, we propose three parameters to quantify the effects of drainage on the flux rope acceleration in the above three aspects, respectively, even in the case that we cannot accurately estimate the parameters $F_n$ and $F_o$ (as it is difficult to determine the boundaries and region of eruptive flux rope). For the first and third aspects, the proposed parameters are $1/M$ and $F_g/M$ (Fig \ref{fig4}d,f) which represent the reciprocal of the mass of upper filament/flux rope and the gravitational acceleration of upper filament/flux rope, respectively. For the second aspect, the proposed parameter is $(\mathrm{d}F_g/\mathrm{d}t)/M$ (Fig \ref{fig4}e), which represents the variation of flux rope acceleration per unit of time due to the change in net force as caused by the change in gravity force, i.e., the rate of change of the flux rope acceleration contributed by the force imbalance due to filament drainage.

It is clear that these three parameters start to quickly increase at $t\approx80$ (Fig \ref{fig4}d-f), indicating that the drainage of filament mass contributes significantly to the acceleration of flux rope since then. In addition, it is worth noting that, for either of these three parameters, the parameter evolutions measured under different mass density thresholds (which are used to identify the filament) are similar (Fig \ref{fig4}d-f), suggesting that the aforementioned results are not qualitatively affected by the uncertainty in identifying the eruptive filament.

The quantitative analyses on the parameter $1/M$ better illustrates the significant contribution of the drainage of filament mass to the flux rope acceleration. The parameter $1/M$ is increased by a factor of 1.7 during $80\le t\le88$ (before slow rise phase) and further by a factor of 2.4 during $88\le t\le95$ (during slow rise phase; Fig \ref{fig4}d). As a comparison, the acceleration of flux rope is increased by a factor of 3.1 during $80\le t\le88$ (before slow rise phase) and by a factor of 5.6 during $88\le t\le95$ (during slow rise phase; Fig \ref{fig2}f). This means that the reduction in the mass of flux rope due to the drainage of filament mass is already able to contribute about a half of the increase in the flux rope acceleration both before and during the slow rise phase, even in the case without considering the change of the net force acting on flux rope.

Lastly, it is interesting to discuss the detailed relationship between the evolution of gravitational acceleration of upper filament ($F_g/M$) and the drainage of filament mass. First, since the parameter $|F_g|/M$ is derived by $\int{\rho |g(z)|\mathrm{d}V}/\int\rho\mathrm{d}V$ ($\mathrm{d}V=\mathrm{d}x\mathrm{d}y\mathrm{d}z$) in the eruptive filament, it could be reduced when the mass density of filament plasmas decreases by a higher proportion at a lower altitude where the parameter $|g(z)|$ is larger. In fact, the drainage of filament mass does occur mainly in the legs (lower part) of eruptive filament rather than its top (Fig \ref{fig4}a-c). This exactly fulfills the aforementioned condition that the mass density of filament plasmas decreases by a higher proportion at a lower altitude, thus allowing the drainage to influence the evolution of parameter $F_g/M$. Second, we note that the parameter $|F_g|/M$ could also be reduced as the gravitational acceleration coefficient in the eruptive filament becomes smaller and smaller while the filament rises to a higher altitude, a process which is not directly related to the drainage. However, as shown in Fig \ref{fig4}i, the average of gravitational acceleration coefficient (derived by $\int{g(z)\mathrm{d}V}/\int\mathrm{d}V$) in the eruptive filament starts to increase quickly at $t\approx86$, later than the onset time of the quick increase of the parameter $F_g/M$ ($t\approx80$). This result indicates that the reduction of the absolute value of gravitational acceleration coefficient ($|g(z)|$) in the filament during its rising is not the main cause of the variation in the evolution of gravitational acceleration ($F_g/M$) of filament around $t=80$. Instead, the main cause should be that the absolute value of the gravity force acting on the upper filament is reduced faster than its mass under the effect of the uneven drainage of filament mass, as mentioned initially.

\subsection{Evolution of Magnetic Reconnection}

\begin{figure*}
\centering
\includegraphics[width=\hsize]{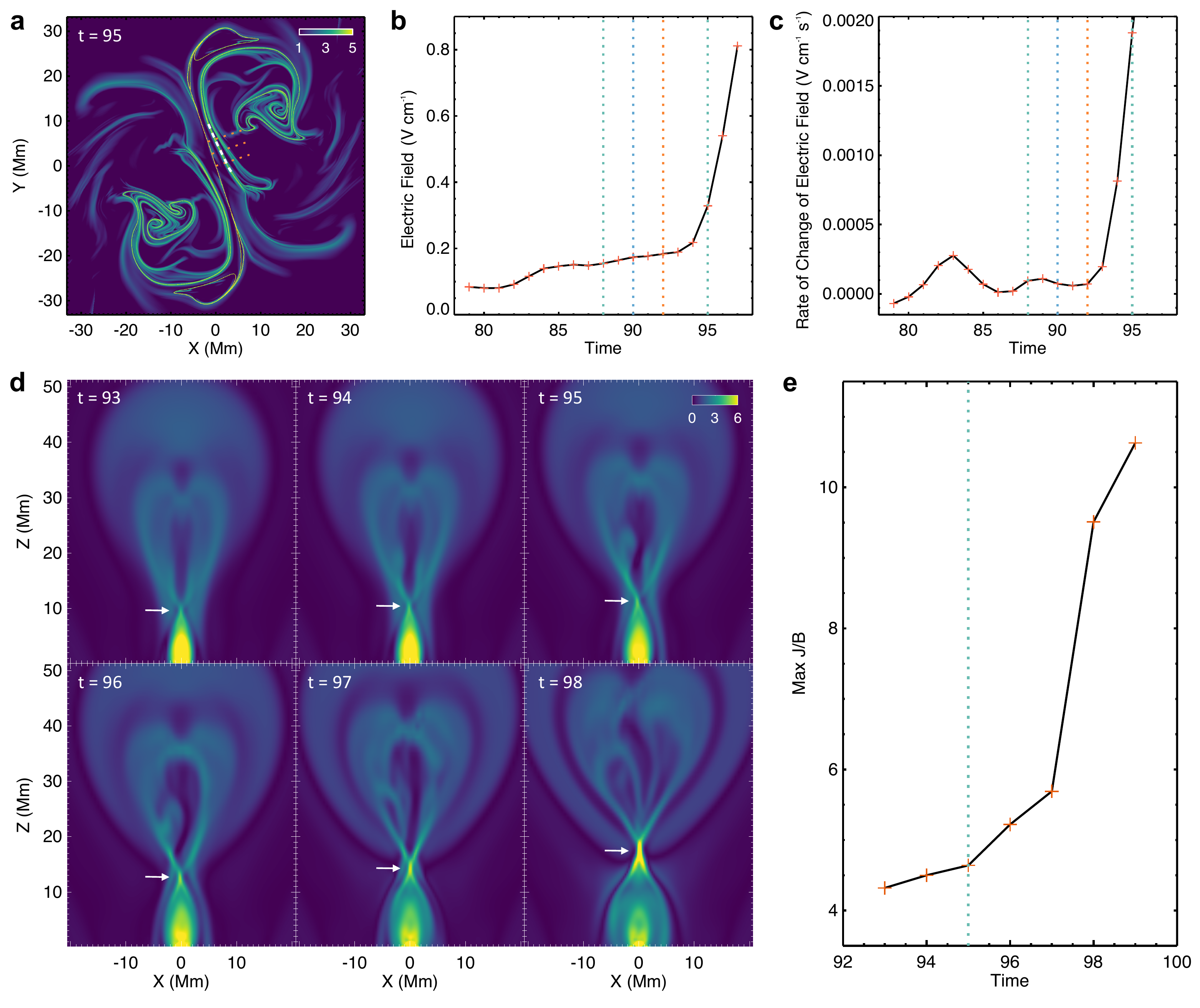}
\caption{\textbf{Analyses on the evolution of magnetic reconnection.} (a) The squashing degree ($\log Q$) map on the cell-center bottom surface at $t=95$. The bright regions represent the QSL footprints, and the white dashed curve marks a section of QSL footprints related to the reconnection region. The three orange dashed lines represent the slits which are used for measuring the separation speed of QSL footprints. (b) The evolution of reconnection electric field in the regions pointed by the arrows in panel d. (c) The rate of change of the reconnection electric field. The green, blue, and orange dashed lines in panels b and c are the same as those in Fig \ref{fig2}. (d) The evolution of $J/B$ in the plane $y=0$. The arrows point to the reconnection regions where the HFT reconnection and the fast magnetic reconnection occur. (e) The evolution of maximum $J/B$ in the reconnection region, the latter of which is pointed by the arrows in panel d. The green dashed line marks the start of the impulsive acceleration phase.}
\label{fig5}
\end{figure*}

As mentioned in Section \ref{split}, an HFT is embedded in the flux rope before the split of flux rope. The HFT can date back to at least $t=79$, as evidenced by its QSL footprints on the cell-center bottom surface. The separation motion of HFT footprints away from the PIL shows that the HFT reconnection has occurred at $t=79$ and contributes to the formation of upper flux rope since then. 

Here, we evaluate the rate of the HFT reconnection and also the fast magnetic reconnection during the eruption with the electric field of magnetic reconnection \citep{Forbes1984}, the latter of which is estimated by a method similar to that in \citetalias{Xing2024}. In simple terms, the reconnection electric field is estimated by the product of the separation speed of the QSL footprints related to the reconnection region (e.g., the HFT) in the direction perpendicular to the PIL on the cell-center bottom surface and the vertical magnetic field strength ($B_z$) at the QSL footprints \citep{Qiu2002}. To calculate the separation speed of QSL footprints, we set three slits perpendicular to the PIL at the cell-center bottom surface (see the orange dashed lines in Fig \ref{fig5}a). For each slit, the separation speed of QSL footprints (marked by the white dashed curve in Fig \ref{fig5}a) is derived by the movement speed of the intersection of the slit and the QSL footprints. We derive one evolution of reconnection electric field from the separation speed and $B_z$ measured along each slit, and the evolution of reconnection electric field displayed in Fig \ref{fig5}b is an average of the three evolutions derived respectively from three slits.

As shown in Fig \ref{fig5}b, the rate of HFT reconnection is almost invariant during $79\le t\le81$. Later, the HFT reconnection has a gradual enhancement during $81<t\le93$ (Fig \ref{fig5}b), with two significant increases of electric field occurring around $t=83$ and $t=89$, respectively (Fig \ref{fig5}c). The magnetic reconnection is sharply strengthened since $t=93$ and until $t=97$ which is the last moment we can measure the electric field (Fig \ref{fig5}b,c).

More specifically, the rate of change of the reconnection electric field has an increase during $87\le t\le89$, a decrease during $89\le t\le91$, and a sharp increase after $t=92$ (Fig \ref{fig5}c). As a comparison, the flux rope acceleration has an increase during $87\le t\le90$, a decrease during $90\le t\le92$, and a sharp increase after $t=92$ (Fig \ref{fig2}f). Therefore, there is a high consistency between the evolutions of magnetic reconnection and flux rope acceleration in these periods, which illustrates the significant effect of the magnetic reconnection on the acceleration of flux rope around the slow rise phase. Actually, \citetalias{Xing2024} has shown that the magnetic reconnection does make a contribution to accelerating the flux rope, by building up the flux rope and pushing the flux rope with its upward reconnection outflows.

We also study the evolution of the current in the reconnection region. One can find that a thin vertical current sheet is formed at $95\le t\le96$ (Fig \ref{fig5}d), which is also evidenced by the rapid increase of maximum $J/B$ in the reconnection region since this moment (Fig \ref{fig5}e). In addition, the reconnection electric field also has an impulsive enhancement during $95\le t\le97$ (Fig \ref{fig5}b). Therefore, we conclude that a fast magnetic reconnection sets in this simulation at $95\le t\le96$.

\subsection{Onset of Torus Instability}

Some evidences suggest that the torus instability \citep{Kliem2006} occurs during the flux rope rising. As shown in Fig \ref{fig5}b, the reconnection electric field starts to sharply increase at $t=93$ which is before the onset of fast magnetic reconnection. The similar phenomenon also appeared in our previous simulation in \citetalias{Xing2024}, where we found the strengthening of HFT reconnection before the start of fast magnetic reconnection and showed that the strengthening is due to the onset of torus instability. Specifically, the torus instability can enhance the HFT reconnection by squeezing the current structure in HFT, as the torus instability drives the flux rope to rise and stretches the overlying field \citep{Janvier2014,Xing2024}. Taking these factors into account, we speculate that the torus instability may set in this simulation around $t=93$, which soon enhances the HFT reconnection (Fig \ref{fig5}b) and accelerates the eruptive flux rope (Fig \ref{fig2}f). This speculation is also supported by the phenomenon that the height of the apex of flux rope axis is almost equal to the distance between the two centers of opposite polarities on the cell-center bottom surface at $t=93$, which also appears in previous CME simulations at their torus instability onset time \citep[e.g.,][]{Aulanier2010,Xing2024}. 

At $t=93$, the flux rope is semicircular (Fig \ref{fig1}) and the decay index of the background potential field at the apex of flux rope axis is 1.76. This decay index is larger than the theoretical critical decay index ($n_c=1.5$) of torus instability in the case of a semicircular flux rope \citep{Kliem2006,Demoulin2010}. However, such a result is reasonable. As shown in Fig 2 of \citet{Jenkins2019}, in the case when the gravity force is taken into account and without mass-draining, the critical height of torus instability onset of a flux rope with mass (marked by the point $E$) is higher than that of a flux rope without mass (marked by the point $C^{\prime}$), which means that the critical decay index of torus instability of the former should be larger than that of the latter. Therefore, the theoretical result \citep{Jenkins2019} supports our simulation result, i.e., that the critical decay index of a flux rope carrying filament mass is larger than those of zero-mass flux ropes \citep{Kliem2006,Demoulin2010}. Actually, in this work, the apex of flux rope axis rises to the altitude where the decay index is equal to 1.5 at as early as $t=83$. However, at this moment, the flux rope is still in the quasi-static phase when the torus instability is unlikely to occur. This further supports that the critical decay index of torus instability in this work should be larger than 1.5.

\section{Initiation Route of Modeled CME}\label{sec3}

Based on the above analyses, the complete initiation route of modeled CME is summarized as follows. After a long-period quasi-static evolution of flux rope, the HFT reconnection and the drainage of filament mass begin to strengthen at $t\approx81$ (Fig \ref{fig4}d-f and Fig \ref{fig5}b). Both of these two mechanisms are continuously enhanced during $81\le t\le88$ (Fig \ref{fig4}d-f and Fig \ref{fig5}b) and make significant contributions to the approach of flux rope towards its initiation in this period. Specifically, the drainage of filament mass contributes at least about a half of the increase of flux rope acceleration during $80\le t\le88$ (see Section \ref{drainage}; Fig \ref{fig2}f and Fig \ref{fig4}d), and the development of the HFT embedded in flux rope sets the stage for the split of flux rope at the start of slow rise phase (see Section \ref{split}; Fig \ref{fig3}). However, the flux rope still rises quasi-statically during $81<t<88$, as these two mechanisms are still relatively weak at their early development stages (Fig \ref{fig4}d-f and Fig \ref{fig5}b).

The split of flux rope/filament starts at $t\approx88$ (Fig \ref{fig3}) and it marks the onset of the CME initiation/slow rise phase. As the HFT is built up, the split of flux rope is quickly developed during $88\le t\le90$ (Fig \ref{fig3}), and the HFT reconnection is also significantly enhanced in this period (Fig \ref{fig5}b,c). Such an enhancement of HFT reconnection leads to the first increase of flux rope acceleration during the slow rise phase ($88\le t\le90$; Fig \ref{fig2}f). In addition, the effects of the drainage of filament mass on the flux rope acceleration are strengthened during $88\le t\le90$ (Fig \ref{fig4}d,f), which also plays a role in causing the slow rise of flux rope.

The acceleration of eruptive flux rope shows a slight decrease during $90\le t\le92$ (Fig \ref{fig2}f). This phenomenon is most likely to be due to the HFT reconnection rather than the drainage of filament mass, as the enhancement of magnetic reconnection slows down (Fig \ref{fig5}c) while the effects of drainage on the flux rope acceleration are enhanced at a faster rate in this period (Fig \ref{fig4}g,h). Again, the response of the evolution of flux rope acceleration to the evolution of HFT reconnection illustrates the important role of magnetic reconnection in driving the slow rise of flux rope.

Later, the acceleration of flux rope increases again during the slow rise phase ($92\le t\le95$; Fig \ref{fig2}f). It is certain that the HFT reconnection is able to contribute to the increase of acceleration, since the reconnection is quickly strengthened during $93\le t\le95$ (Fig \ref{fig5}b). The drainage of filament mass can also make an important contribution to the increase of flux rope acceleration, considering that it is also enhanced in this period (Fig \ref{fig4}d-f). Actually, even if the change in the net force acting on the flux rope is not taken into account, the reduction of the mass of filament plasmas is able to result in a half of the increase in the flux rope acceleration during the slow rise phase (see Section \ref{drainage}). In addition, the torus instability, which may start at $t=93$, can also directly accelerate the flux rope by the hoop force since then. Finally, the fast magnetic reconnection starts at $95\le t\le96$ (Fig \ref{fig5}) and it marks the end of the CME initiation/slow rise phase (Fig \ref{fig2}c).

\section{Summary and Discussions}\label{sec4}
In this work, we successfully realize the formation and eruption of a CME containing filament mass by performing a state-of-the-art 3D full-MHD simulation. The filament is naturally formed during the growth of the pre-eruptive flux rope, the latter of which is built up in the flux cancellation process. Both of the filament and the flux rope supporting it are split into two parts shortly before the eruption, thus in a configuration of the so-called double-decker filaments/flux ropes \citep{Gibson2006,Liu2012,Pan2021}. The motion of the upper flux rope containing the eruptive filament reproduces the slow rise of filaments shortly before the impulsive injection as discovered in observations \citep{Sterling2011,Cheng2020}.

We reveal a complete initiation route of CMEs which carry dense filament plasmas by comprehensively analyzing the simulation. The initiation (slow rise) of flux rope is triggered by the strengthening HFT reconnection which builds up the flux rope and the enhanced drainage of filament mass in the flux rope. With the development of the HFT, the split of flux rope/filament starts, indicating the start of CME initiation/slow rise phase. Then, the HFT reconnection and the drainage of filament mass are further enhanced, driving the slow rise of flux rope and pushing the flux rope to the critical altitude of the onset of torus instability. The torus instability breaks the stable equilibria of flux rope and continues to drive its slow rise together with the HFT reconnection and the filament drainage. As the flux rope slowly rises, the current system in the reconnection region gradually grows and finally evolves into a thin and long current sheet. The fast magnetic reconnection then starts in the current sheet, marking the end of the initiation process/slow rise phase.

The role of the filament mass in the CME initiation is highlighted through a comparison between the two CME simulations in \citetalias{Xing2024} and this work. It is clear that the initiation route of a flux rope with coronal mass density \citepalias{Xing2024} and that of a flux rope carrying filament mass (this work) are similar in many aspects. In both cases, the magnetic reconnection occurring in the HFT below the eruptive flux rope plays a role in triggering and driving the initiation/slow rise of flux rope, and the torus instability starting during the initiation process also makes a contribution to driving the initiation/slow rise. Nevertheless, these two simulations still show significant differences. In \citetalias{Xing2024}, we did not find a significant drainage of flux-rope plasmas and thus the drainage is not considered as a mechanism of initiation. The absent of significant drainage there is due to that (1) a coronal-mass-density flux rope does not have much plasmas to unload, and (2) the gravity force acting on the flux rope, which is much less than the Lorentz force and thermal pressure gradient force acting on the flux rope (Fig 6 in \citetalias{Xing2024}), makes the drainage not easy to occur. On the contrary, in this work, the presence of filament in the eruptive flux rope makes a significant drainage of flux-rope plasmas a reality (Fig \ref{fig4}a-c) by providing sufficient dense plasmas to unload and a strong gravity force against other forces (see Appendix \ref{app2} and Fig \ref{figA1}). As a consequence, the drainage of filament mass is found to play an important role in accelerating the flux rope before and during the slow rise phase since the onset of its enhancement (Fig \ref{fig4}d-f). In summary, the filament mass has a significant effect on triggering and driving the CME initiation through promoting the drainage of plasmas in the flux rope.

It is also worth comparing our simulation with other simulations of filament eruptions \citep{Fan2017,Fan2018,Fan2019,Fan2020}. Similar to our simulation, their series of work achieved the self-consistent formation and eruption of filaments and the drainage of filament mass always occurred prior to the eruption onset, although their filaments were formed under the condensation model unlike our results \citep{Fan2017,Fan2018,Fan2019,Fan2020}. Our finding that the flux rope containing filament mass has a higher critical decay index of the torus instability than that of zero-mass flux rope is also in agreement with their result that the filament mass had a role in suppressing the onset of ideal MHD instability \citep{Fan2018,Fan2020}. In addition, \citet{Fan2020} showed that the drainage of filament mass could cause the flux rope to lose its equilibrium earlier through some preliminary explorations. Compared to their works, our work makes more detailed analyses on the CME initiation/slow rise phase by more precisely relating the kinematics of flux rope with the physical processes involved in the flux rope evolution. On this basis, as a new result, our work determines the explicit role of the drainage of filament mass and HFT reconnection in triggering and driving the initiation/slow rise of flux ropes containing filament mass.

The initiation route of CMEs containing filament mass revealed in this work is also supported by the observations of filament eruption. On the one hand, the filament mass is often observed to be drained before the filament eruption, suggesting that the mass unloading plays a role in promoting the onset of eruption \citep{Seaton2011,Bi2014,Huang2014,Jenkins2018,Filippov2020,Dai2021}. This is in agreement with our result that the drainage of filament mass triggers and drives the slow rise of flux rope towards the eruption onset. On the other hand, an apparent observational feature of the HFT reconnection below the eruptive filament, that is, the split of filament, is also often observed in a period shortly before the eruption \citep{Cheng2018,Zheng2019,Chen2021,Pan2021,Hou2023,Sun2023}. Specifically, each of \citet{Cheng2018} and \citet{Pan2021} reported a filament eruption event and showed the kinematic evolution of eruptive filament. For each of these two events, the split of filament occurs over ten minutes before the eruption. Between the split and the eruption onset, the eruptive filament exhibits a slow rise with a velocity of tens of km s$^{-1}$ and an acceleration of about tens of m s$^{-2}$. These observations support our result that the HFT reconnection plays a role in triggering and driving the initiation of flux ropes containing filament mass.

Lastly, as a summary of \citetalias{Xing2024} and this work, we propose here the precursors of CME initiation in observations for a better prediction of solar eruptions. For both coronal-mass-density flux ropes appearing as hot channels and flux ropes carrying filament mass, the precursors of their initiation could be the following phenomena associated with the HFT reconnection: the pre-flare loops and X-ray emissions \citep{Cheng2023}, and the split of pre-eruptive structure. In addition, for flux ropes carrying filament mass, the precursors of their initiation also include an enhanced drainage of filament mass. These precursor features are highly worth to be examined by more advanced observations and state-of-the-art numerical simulations in the future.

\section*{acknowledgments}
We thank the referee for reviewing the manuscript and providing valuable comments and suggestions.
We thank Jinhan Guo and Can Wang for valuable discussions.
C.X. and X.C. acknowledge the supports by the NSFC under grant 12403066, the Jiangsu NSF under grant BK20241187, the Fundamental Research Funds for the Central Universities under grant 2024300348, the Specialized Research Fund for State Key Laboratories, the Postdoctoral Fellowship Program of CPSF under grant GZC20240693, and the Jiangsu Funding Program for Excellent Postdoctoral Talent.
G.A. acknowledges financial support from the Programme National Soleil Terre (PNST) of the CNRS/INSU also co-funded by CNES and CEA.
The numerical calculation in this work was done on the computing facilities in High Performance Computing Center of Nanjing University.
MPI-AMRVAC 2.2, the code used to perform the simulation in this paper is an open source code. It is available from the website \url{https://amrvac.org/index.html}. The code used to calculate the squashing degree is available from the website \url{https://github.com/el2718/FastQSL}.

\clearpage
\newpage
\appendix
\beginsupplement

\section{3D Full-MHD Simulation of a CME Event Containing Filament Mass}\label{app1}
\subsection{Equations and Numerical Methods}
We achieve a three-dimensional (3D) observationally-inspired full-MHD simulation of the formation and eruption process of a filament with the code MPI-AMRVAC \citep{Xia2018}. The simulation solves the following equations in Cartesian coordinates:
\begin{equation}\label{eq1}
\frac{\partial\rho}{\partial t}+\nabla\cdot(\rho\boldsymbol{v}) = 0
\end{equation}
\begin{equation}\label{eq2}
\begin{aligned}
\frac{\partial(\rho\boldsymbol{v})}{\partial t}+\nabla\cdot[\rho\boldsymbol{v}\boldsymbol{v}+(p+\frac{\boldsymbol{B}^2}{2\mu_0})\boldsymbol{I}-\frac{\boldsymbol{B}\boldsymbol{B}}{\mu_0}] = \rho\boldsymbol{g}+2\mu\nabla\cdot[\boldsymbol{S}-\frac{1}{3}(\nabla\cdot\boldsymbol{v})\boldsymbol{I}]
\end{aligned}
\end{equation}
\begin{equation}\label{eq3}
\frac{\partial \boldsymbol{B}}{\partial t}+\nabla\cdot(\boldsymbol{v}\boldsymbol{B}-\boldsymbol{B}\boldsymbol{v}+\psi\boldsymbol{I}) = -\nabla\times(\eta\boldsymbol{J})
\end{equation}
\begin{equation}\label{eq4}
\begin{aligned}
\frac{\partial e_{\textup{int}}}{\partial t}+\nabla\cdot(e_{\textup{int}}\boldsymbol{v})=-p\nabla\cdot\boldsymbol{v}+2\mu[\boldsymbol{S}:\boldsymbol{S}-\frac{1}{3}(\nabla\cdot\boldsymbol{v})^2]+\eta J^2+\nabla\cdot[\kappa_{||}(\boldsymbol{b}\cdot\nabla T)\boldsymbol{b}]+H-R
\end{aligned}
\end{equation}
\begin{equation}\label{eq5}
\nabla\times\boldsymbol{B} = \mu_0\boldsymbol{J}
\end{equation}
\begin{equation}\label{eq6}
\frac{\partial\psi}{\partial t}+c_h^2\nabla\cdot\boldsymbol{B} = -\frac{c_h^2}{c_p^2}\psi.
\end{equation}
Here, $\rho$, $p$, $T$, $e_{\textup{int}}$, $\boldsymbol{v}$, $\boldsymbol{B}$, $\boldsymbol{J}$ represent the mass density, thermal pressure, temperature, internal energy, velocity, magnetic field and current density, respectively. All parameters in Equations \ref{eq1}--\ref{eq6} except $H$ and $R$ have the same meanings as those in Equations 1--6 of \citetalias{Xing2024}. The parameter $H=H_0B^ar^b\rho^c$ represents the volumetric background heating rate \citep{Mok2016}; $B$, $r$, and $\rho$ represent the magnetic field strength, local curvature radius of magnetic field line, and mass density, respectively; $H_0$, $a$, $b$, and $c$ are set to $10^{-4}$ erg cm$^{-3}$ s$^{-1}$, 1.75, 0.75, and 0.125, respectively. The parameter $R$ represents the optically thin radiative loss with the \textit{SPEX} cooling curve \citep{Schure2009}. The equations are solved dimensionlessly, and the units to dimensionalize the length, time, mass density, thermal pressure, temperature, velocity, and magnetic field strength are 10 Mm, 85.87 s, $2.34\times10^{-15}$ g cm$^{-3}$, 0.32 erg cm$^{-3}$, 1 MK, 116.45 km s$^{-1}$, and 2 G, respectively (in this paper, parameters are dimensionless unless indicated).

The simulation domain is a cube of $-7\le x\le7$, $-7\le y\le7$, and $0\le z\le14$. The domain is resolved by 180 symmetric-stretched grids with a stretched ratio of 1.019 in both $x$ and $y$ directions and by 140 unidirectional-stretched grids with a stretched ratio of 1.022 in $z$ direction. Benefitting from the stretched grids, the finest resolutions of domain are about 300 km in $x$ and $y$ directions and about 150 km in $z$ direction. To achieve an accurate and stable simulation, we use the HLL scheme, the third-step Runge-Kutta time discretization method, the fifth-order weno5-limited reconstruction \citep{Jiang1996,Shu2009}, the generalized Lagrange multiplier (GLM) $\nabla\cdot\boldsymbol{B}$ cleaning method \citep{Dedner2002}, and the magnetic field splitting method \citep{Tanaka1994,Xia2018}.

\subsection{Initial and Boundary Conditions}
The initial atmosphere of this simulation is a stratified fully-ionized atmosphere including the chromosphere, transition region, and corona. The temperature of the initial atmosphere is set to:
\begin{equation}\label{eq7}
T(z) = \left\{
\begin{aligned}
T_{ch}+\frac{1}{2}(T_{co}-T_{ch})(\tanh(\frac{z-h_{tr}-0.027}{w_{tr}})+1) & & & & & z\le h_{tr} \\
(\frac{7}{2}\frac{F_c}{\kappa}(z-h_{tr})+T_{tr}^{7/2})^{2/7} & & & & & z> h_{tr},
\end{aligned}
\right.
\end{equation}
where $T_{ch}=8\times10^{-3}$, $T_{tr}=0.16$, $T_{co}=1.5$, $h_{tr}=0.243$, $w_{tr}=0.02$, $F_c=0.054$, and $\kappa=0.22$. The gravitational acceleration coefficient is set to:
\begin{equation}\label{eq8}
g(z)=g_0R_{sun}^2/(R_{sun}+z)^2<0,
\end{equation}
where $g_0=-0.20$ and $R_{sun}=69.55$. The mass density and thermal pressure of the initial atmosphere are then derived under the hydrostatic assumption and with $\rho_{z=0}=2.1\times10^{4}$ as input. It is worth noting that the initial atmosphere, although hydrostatic, still needs to be relaxed during $0\le t\le12$ to reach a thermal equilibrium under the background heating, thermal conduction, and radiative cooling.

The configuration of the initial magnetic field is the same as that in \citetalias{Xing2024}, while the magnetic field strength is stronger in this simulation by setting $c_1=240$, $c_2=-240$, $c_3=180$, and $c_4=-180$ (see more details of these parameters in \citetalias{Xing2024}). The magnetic field strength in the plane $z=0$ is thus up to about 360 G. The initial velocity and the initial parameter $\psi$ (used in the GLM method) are set to zero in the whole physical domain.

The boundary conditions of $\vec{v}$, $\vec{B}$, and $\psi$ are the same as those in \citetalias{Xing2024}. For the mass density and thermal pressure, they are fixed to their initial values  at the bottom boundary, and are derived by the second-order zero-gradient extrapolation at the other five boundaries.

\subsection{Line-tied Driving Motions}
Following the ideas in \citetalias{Xing2024}, here we still impose the line-tied shearing and converging driving motions on the bottom to drive the magnetic field and form the flux rope. During $12<t\le30$, we impose a shearing motion ($v_x^s, v_y^s, v_z^s$) on the first layer of the physical domain (labeled as the layer $k=1$) to drive the initial field to a highly sheared state:
\begin{equation}\label{eq9}
\begin{gathered}
v_x^s(k=1; t)=\gamma(t)v_0^{max}\Psi_0(t)\partial_y\Psi(t) \\
v_y^s(k=1; t)=-\gamma(t)v_0^{max}\Psi_0(t)\partial_x\Psi(t) \\
v_z^s(k=1; t)=0 \\
\gamma(t)=\left\{
\begin{aligned}
\frac{1}{2}\tanh[3.75(t-13)]+\frac{1}{2} & & & & & & 12<t<28 \\
-\frac{1}{2}\tanh[3.75(t-29)]+\frac{1}{2} & & & & & & 28\le t\le30.
\end{aligned}
\right.
\end{gathered}
\end{equation}
During $30<t\le101$, we impose a converging motion ($v_x^c, v_y^c, v_z^c$) on the same layer to promote the magnetic reconnection along the PIL:
\begin{equation}\label{eq10}
\begin{gathered}
v_x^c(k=1; t)=\gamma(t)v_0^{max}\Psi_0(t)\partial_x\Psi(t) \\
v_y^c(k=1; t)=\gamma(t)v_0^{max}\Psi_0(t)\partial_y\Psi(t) \\
v_z^c(k=1; t)=0 \\
\gamma(t)=\frac{1}{2}\tanh[3.75(t-31)]+\frac{1}{2} \ \ \ \ \ \ 30<t\le101.
\end{gathered}
\end{equation}
For both the shearing and converging motions, they have:
\begin{equation}\label{eq11}
\Psi(t)=\exp[-\Psi_1(\frac{B_z(k=1; t)}{B_z^{max}(k=1; t)})^2],
\end{equation}
where $\Psi_1=5.5$ for the shearing motion and $\Psi_1=0.5$ for the converging motion. The maximum speed of the driving motions $v_0^{max}$ is set to 18.63 km s$^{-1}$ during $12<t\le30$ and 4.66 km s$^{-1}$ during $30<t\le101$.

The setups to achieve the line-tied bottom boundary condition are the same as those in \citetalias{Xing2024}. The only change is about the resistivity: during the shearing phase ($12<t\le30$), the resistivity $\eta$ is set to zero at the layer $k=1$ and to be uniform ($\eta=10^{-4}$) in the whole simulation domain except the layer $k=1$; during the converging phase ($30<t\le101$), the resistivity $\eta$ is set to be uniform ($\eta=2\times10^{-4}$) in the whole domain.

\section{Dynamic Analysis of Eruptive Flux Rope and Filament}\label{app2}

\begin{figure*}
\centering
\includegraphics[width=\hsize]{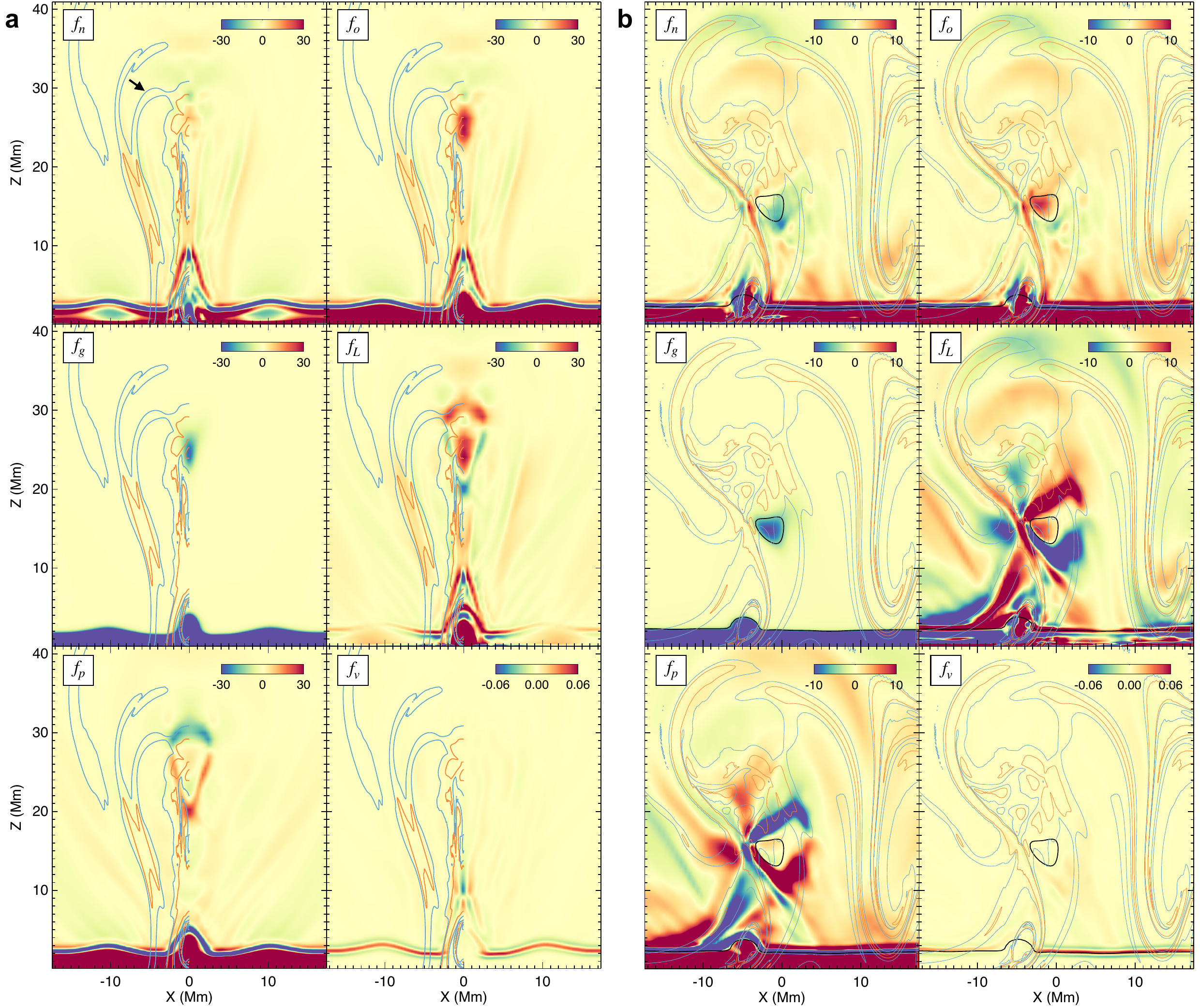}
\caption{\textbf{Distributions of forces in the planes $y=0$ and $y=2$ at $t=92$.} (a) From left to right and top to bottom, they are the $z$-components of the net force ($f_n$), the net of other forces in addition to the gravity force ($f_o$), the gravity force ($f_g$), the Lorentz force ($f_L$), the thermal pressure gradient force ($f_p$), and the viscous force ($f_v$) in the plane $y=0$. In each sub-panel, the blue and orange curves represent the contours of $\textup{log}Q=2$ and $\textup{log}Q=3$, respectively, in the domain $x<0$. In the first sub-panel, the black arrow points to the contours of $\textup{log}Q=2$ which roughly outline the eruptive flux rope. (b) Similar to panel a but showing the distributions of forces in the plane $y=2$. The blue and orange curves represent the contours of $\textup{log}Q=2$ and $\textup{log}Q=3$, respectively. The black curves represent the contours of $f_g=-5$.}
\label{figA1}
\end{figure*}

In addition to studying the integral gravity force ($F_g$) acting on the eruptive flux rope, we also analyze the distributions of forces within the eruptive flux rope to better understand its dynamics during the slow rise phase. To be specific, we calculate the distributions of the $z$-direction components of a series of forces, including the gravity force ($f_g=\rho g(z)$), the Lorentz force ($f_L=(\boldsymbol{J}\times\boldsymbol{B})_z$), the thermal pressure gradient force ($f_p=-\nabla\cdot(p\boldsymbol{e_z})$), and the viscous force ($f_v=(2\mu\nabla\cdot[\boldsymbol{S}-\frac{1}{3}(\nabla\cdot\boldsymbol{v})\boldsymbol{I}])_z$). We also calculate the distributions of the $z$-direction component of the net force, $f_n=f_g+f_L+f_p+f_v$, and the $z$-direction component of the net of other forces in addition to the gravity force, $f_o=f_L+f_p+f_v$.

Fig \ref{figA1}a exhibits the distributions of these forces in the plane $y=0$ which is perpendicular to the axis of the eruptive flux rope at $t=92$ (at the middle of the slow rise phase). The contours of $\textup{log}Q=2$ (pointed by the black arrow) roughly outline the eruptive flux rope which lies between $z\approx10$ Mm and $z\approx31$ Mm. The eruptive filament, as indicated by the region with a strong downward gravity force, is located around the altitude $z=25$ Mm. In addition, Fig \ref{figA1}b shows the distributions of forces in the plane $y=2$ which cuts a leg of filament (Fig \ref{fig4}b) at $t=92$. In this plane, the eruptive filament, also indicated by the region with a strong gravity force (outlined by a black contour), is located around the altitude $z=15$ Mm. However, it is difficult to identify the cross section of eruptive flux rope with the contours of $\textup{log}Q$ in this plane.

As shown in Fig \ref{figA1}a, it is clear that the gravity force, the Lorentz force and the thermal pressure gradient force make important contributions to driving the eruptive flux rope, while the viscous force plays a less important role. In particular, within the eruptive flux rope, the strong downward gravity force is concentrated at the eruptive filament (Fig \ref{figA1}a) and it is comparable to the net of other forces ($f_o$) there in magnitude (Fig \ref{figA1}a,b), the latter of which is mainly contributed by the upward Lorentz force (Fig \ref{figA1}a,b). Due to the combined effect of the gravity force and the other forces, the eruptive filament is subjected to an upward net force ($f_n$) at its top (Fig \ref{figA1}a) and a downward net force in its leg (Fig \ref{figA1}b), which well explains the upward flow at the top of filament and the downward flow (i.e, the drainage of filament mass) in the filament leg (Fig \ref{fig4}b). These results clearly show that, in this simulation, (1) it is the gravity force, against the Lorentz force, that leads to the drainage of filament mass in the leg of filament; (2) the gravity force has the ability to significantly affect the dynamics and thus the kinematics of the eruptive flux rope, which is obviously different from the situation in \citetalias{Xing2024}.

\end{document}